\documentclass[aps,prb,twocolumn,showpacs,preprintnumbers,amsmath,amssymb,superscriptaddress]{revtex4}%

\usepackage{graphicx}%
\usepackage{dcolumn}
\usepackage{amsmath}
\usepackage{color}
\usepackage{multirow}

\begin{document}

\title{In-plane magnetic penetration depth of superconducting
CaKFe$_4$As$_4$}
\author{Rustem~Khasanov}
 \email{rustem.khasanov@psi.ch}
 \affiliation{Laboratory for Muon Spin Spectroscopy, Paul Scherrer Institut, CH-5232 Villigen PSI, Switzerland}
\author{William R. Meier}
 \affiliation{Division of Materials Science and Engineering, Ames Laboratory, Ames, Iowa 50011, USA}
 \affiliation{Department of Physics and Astronomy, Iowa State University, Ames, Iowa 50011, USA}
\author{Yun Wu}
 \affiliation{Division of Materials Science and Engineering, Ames Laboratory, Ames, Iowa 50011, USA}
 \affiliation{Department of Physics and Astronomy, Iowa State University, Ames, Iowa 50011, USA}
\author{Daixiang Mou}
 \affiliation{Division of Materials Science and Engineering, Ames Laboratory, Ames, Iowa 50011, USA}
 \affiliation{Department of Physics and Astronomy, Iowa State University, Ames, Iowa 50011, USA}
\author{Sergey L. Bud'ko}
 \affiliation{Division of Materials Science and Engineering, Ames Laboratory, Ames, Iowa 50011, USA}
 \affiliation{Department of Physics and Astronomy, Iowa State University, Ames, Iowa 50011, USA}
\author{Ilya Eremin}
 \affiliation{Institut fur Theoretische Physik III, Ruhr-Universitat Bochum, 44801 Bochum, Germany}
\author{Hubertus Luetkens}
 \affiliation{Laboratory for Muon Spin Spectroscopy, Paul Scherrer Institut, CH-5232 Villigen PSI, Switzerland}
\author{Adam Kaminski}
 \affiliation{Division of Materials Science and Engineering, Ames Laboratory, Ames, Iowa 50011, USA}
 \affiliation{Department of Physics and Astronomy, Iowa State University, Ames, Iowa 50011, USA}
\author{Paul C. Canfield}
 \affiliation{Division of Materials Science and Engineering, Ames Laboratory, Ames, Iowa 50011, USA}
 \affiliation{Department of Physics and Astronomy, Iowa State University, Ames, Iowa 50011, USA}
\author{Alex Amato}
 \affiliation{Laboratory for Muon Spin Spectroscopy, Paul Scherrer Institut, CH-5232 Villigen PSI, Switzerland}

\begin{abstract}
The temperature dependence of the in-plane magnetic penetration depth ($\lambda_{ab}$) in an extensively characterized sample of superconducting CaKFe$_4$As$_4$ ($T_{\rm c}\simeq35$~K) was investigated using muon-spin rotation ($\mu$SR). A comparison of $\lambda_{ab}^{-2}(T)$ measured by $\mu$SR with the one inferred from ARPES data confirms the presence of multiple gaps at the Fermi level. An agreement between $\mu$SR and ARPES requires the presence of additional bands, which are not resolved by ARPES experiments. These bands are characterised by small supercondcting gaps with an average zero-temperature value of $\Delta_{0} =$~2.4(2)~meV.  Our data suggest that in CaKFe$_4$As$_4$ the $s^\pm$ order parameter symmetry acquires a more sophisticated form by allowing a sign change not only between electron and hole pockets, but also within pockets of similar type.
\end{abstract}


\maketitle

Since their discovery, iron based superconductors (Fe-SC's) have attracted much interest. This broad class of materials exhibits unconventional superconducting properties, due to the strong interplay of superconductivity with various electronic ground states, including a nematic phase and spin-density wave magnetism.\cite{Kamihara_JAPS_2008, Hsu_PNAS_2008, Stewart_RMP_2011, Chen_NSR_2014, Paglione_NatPh_2010}
Recently a new Fe-SC family was synthesized, namely ${\it AeA}$Fe$_4$As$_4$ ({\it Ae} = Ca, Sr, Eu and $A$ = K, Rb, Cs), with superconducting transition temperature ($T_{\rm c}$) reaching $\simeq 36$~K.\cite{Iyo_JAPS_2016, Meier_PRB_2016} CaKFe$_4$As$_4$ (CaK1144) is currently the most studied representative of the {\it AeA}1144 Fe-SC family
with $T_{\rm c}\simeq 35$~K and an estimated upper critical field $H_{\rm c2}^{\perp c}\simeq 92$~T.\cite{Iyo_JAPS_2016,Meier_PRB_2016}
High-resolution angular resolved photoemission (ARPES), nuclear magnetic resonance, tunneling and penetration depth measurements, \cite{Mou_PRL_2016, Cui_PRB_2017, Cho_PRB_2017,Fente_Arxiv_2016, Biswas_PRB_2017} as well as density functional theory (DFT) calculations \cite{Mou_PRL_2016, Lochner_PRB_2017} support multiband superconductivity in CaKFe$_4$As$_4$ with Cooper pairing occurring in electron- and hole-like bands.

In spite of the good agreement between results obtained by different techniques, the gap structure as well as the gap magnitudes are still not conclusively determined. DFT calculations suggest that ten bands cross the Fermi level (6 hole- and 4 electron-like bands).\cite{Mou_PRL_2016, Lochner_PRB_2017} ARPES experiments, on the other hand, reveal only the presence of four bands ($\alpha-$, $\beta-$, $\gamma-$, and $\delta-$bands) with nearly isotropic superconducting energy gaps, with the corresponding zero-temperature values of $\Delta_{0,\alpha}=10.5$, $\Delta_{0,\beta}=13$, $\Delta_{0,\delta}=8$, and $\Delta_{0,\gamma}=12$~meV.\cite{Mou_PRL_2016}
Tunneling experiments show that the superconducting gaps are spread between 1 and 10 meV with a broad peak appearing at around 3~meV, while the superfluid density measurements suggest nodeless two-gap superconductivity  with a larger gap of $\sim 6-10$~meV and a smaller gap of $\sim 1-4$~meV.\cite{Cho_PRB_2017,Fente_Arxiv_2016,Biswas_PRB_2017}

In this paper, we report on measurements of the in-plane magnetic penetration depth ($\lambda_{ab}$) and the vortex core size ($\xi_{ab}$) in a CaKFe$_4$As$_4$ single crystal sample by means of the muon-spin rotation ($\mu$SR) technique. The obtained temperature dependence of $\lambda_{ab}$ was compared with the calculations based on the analysis of the electronic band structure and the momentum dependent superconducting gap extracted from the ARPES data. An agreement between the results obtained by both techniques requires the presence of additional bands  which were not resolved in ARPES experiments.

CaKFe$_4$As$_4$ single crystals were grown from a high-temperature Fe-As rich melt and extensively characterized via thermodynamic and transport measurements.\cite{Meier_PRB_2016} A fraction of the ARPES data was previously reported in Ref.~\onlinecite{Mou_PRL_2016}. A crystal with dimensions of $\simeq 4.0 \; \times \; 4.0 \; \times \; 0.1$~mm$^3$ was used for the $\mu$SR experiments, which  were carried out at the $\pi$M3 beam line using the GPS spectrometer (Paul Scherrer Institute, Switzerland).\cite{Amato_RSI_2017} Transverse-field (TF) $\mu$SR measurements were performed at temperatures from $\simeq$1.5 to 50~K. The external magnetic field ($H_{\rm ap}$) ranging from 5 to 580~mT was applied along the crystallographic $c-$axis of the crystal. A spin rotator was used to orient the initial spin polarization of the muon beam at $45^{\rm o}$ degrees with respect to the applied field. A special sample holder  designed to measure thin samples by means of $\mu$SR was used.\cite{Khasanov_PRB_2016} The experimental data were analyzed using the MUSRFIT package.\cite{MUSRFIT}

\begin{figure}[htb]
\includegraphics[width=0.85\linewidth]{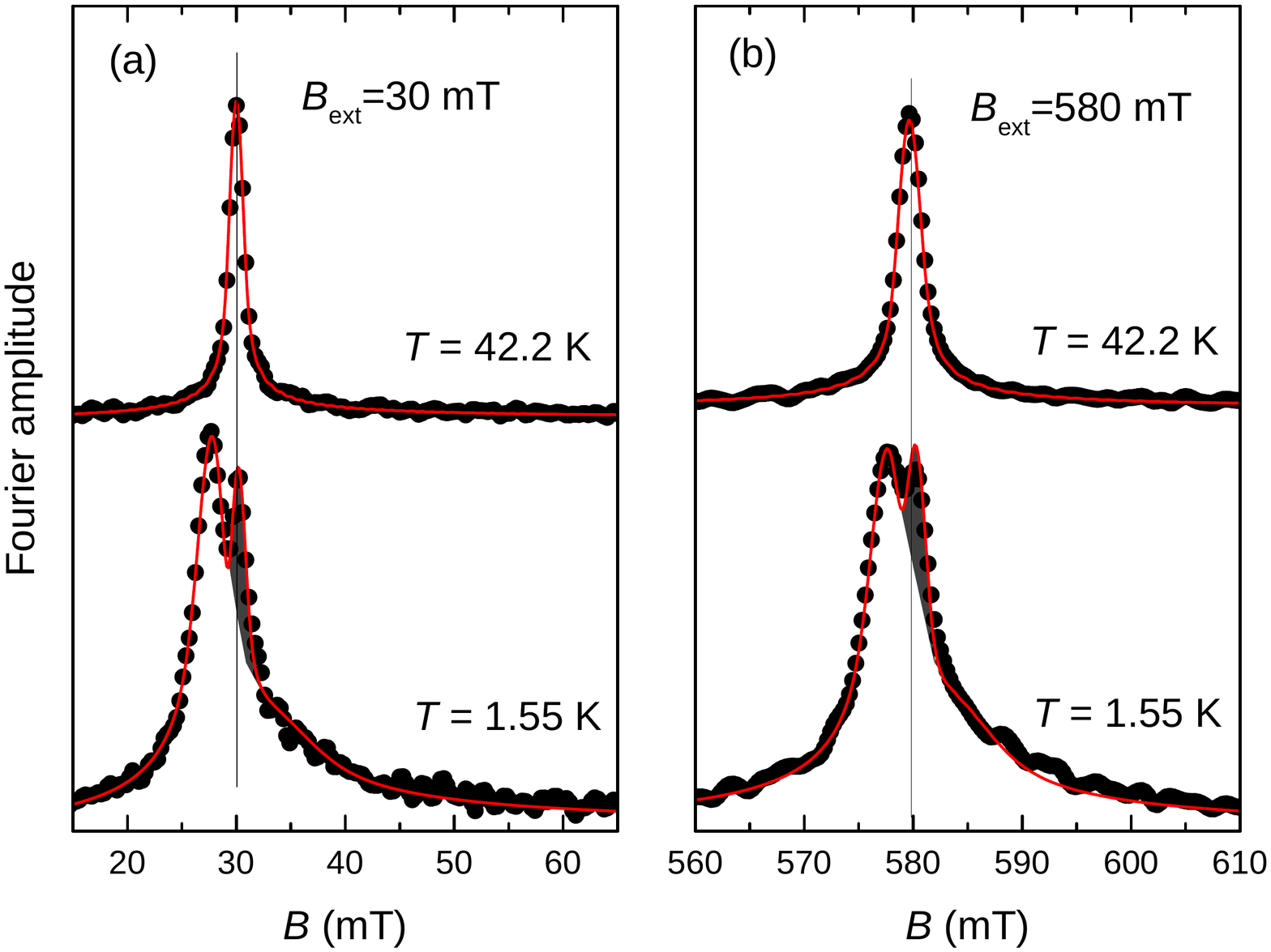}
%
\caption{Fourier transform of the TF-$\mu$SR time spectra obtained in an applied field of $\mu_0H_{\rm ap}=30$~mT (a)  and $\mu_0H_{\rm ap}=580$~mT (b) above ($T=42.2$~K) and below ($T=1.55$~K) $T_{\rm c}\simeq 35$~K. Red lines are the two ($T=42.2$~K) and three ($T=1.55$~K) component Gaussian fits corresponding to the field distribution $P(B)$ described by Eq.~\ref{eq:PB}. The peak at $B=\mu_0H_{\rm ap}$  represents the background signal.   }
 \label{fig:Fourier-transform}
\end{figure}

Figure \ref{fig:Fourier-transform} shows the Fourier transform of TF-$\mu$SR time spectra reflecting the internal field distribution $P(B)$ in the CaKFe$_4$As$_4$ single crystal sample.  The results of field-cooled measurements in a field of $\mu_0 H_{\rm ap}=30$~mT (panel a) and 580~mT (panel b) above ($T = 42.2$~K) and below ($T = 1.55$~K) $T_{\rm c}\simeq 35$~K are presented. The asymmetric $P(B)$ distributions at $T=1.55$~K possess the basic features expected for an ordered vortex lattice, namely: the cutoff at low fields, the peak shifted below $H_{\rm ap}$, and the long tail towards the high field direction (see {\it e.g.} Refs.~\onlinecite{Maisuradze_JPCM_2009, Khasanov_PRB_2016} and references therein). The sharp peak at $B=\mu_0H_{\rm ap}$ represents the residual background signal from muons missing the sample.
The $\mu$SR time spectra  were analyzed using a three component Gaussian expression  with the first (the temperature and field independent) component corresponding to the background contribution, and another two components accounting for the asymmetric $P(B)$ distribution in the mixed state of superconductor (see the Supplemental Information, Ref.~\onlinecite{Supplemental_Material}, and Refs.~\onlinecite{Khasanov_PRB_2006, Khasanov_PRL_2007} for details). The time-domain expression is equivalent to a distribution in the field-domain:
\begin{eqnarray}
P(B)&=&P(B)_{b}+P(B)_{s} =
\frac{\gamma_\mu A_{b}}{\sigma_{b}} \exp \left(-\frac{\gamma_\mu^2(B-B_{b})^2}{2 \sigma_{b}^2} \right)  \nonumber \\
&&+\sum_{i=1}^2 \frac{\gamma_\mu A_i}{\sigma_i} \exp \left(-\frac{\gamma_\mu^2(B-B_i)^2}{2 \sigma_i^2} \right).
\label{eq:PB}
\end{eqnarray}
Indexes $b$ and $s$ correspond to the background and the sample contributions, respectively. $A_b$($A_{i}$), $\sigma_b$($\sigma_i$), and $B_b$($B_i$) are the asymmetry,  the relaxation rate, and the mean field of the background ($i-$th sample)  component. $\gamma_\mu = 2\pi\times135.5342$~MHz/T is the muon gyromagnetic ratio. For $T\geq T_{\rm c}$ the analysis of the sample contribution was simplified to a single Gaussian lineshape.
\begin{figure}[htb]
\includegraphics[width=1.0\linewidth]{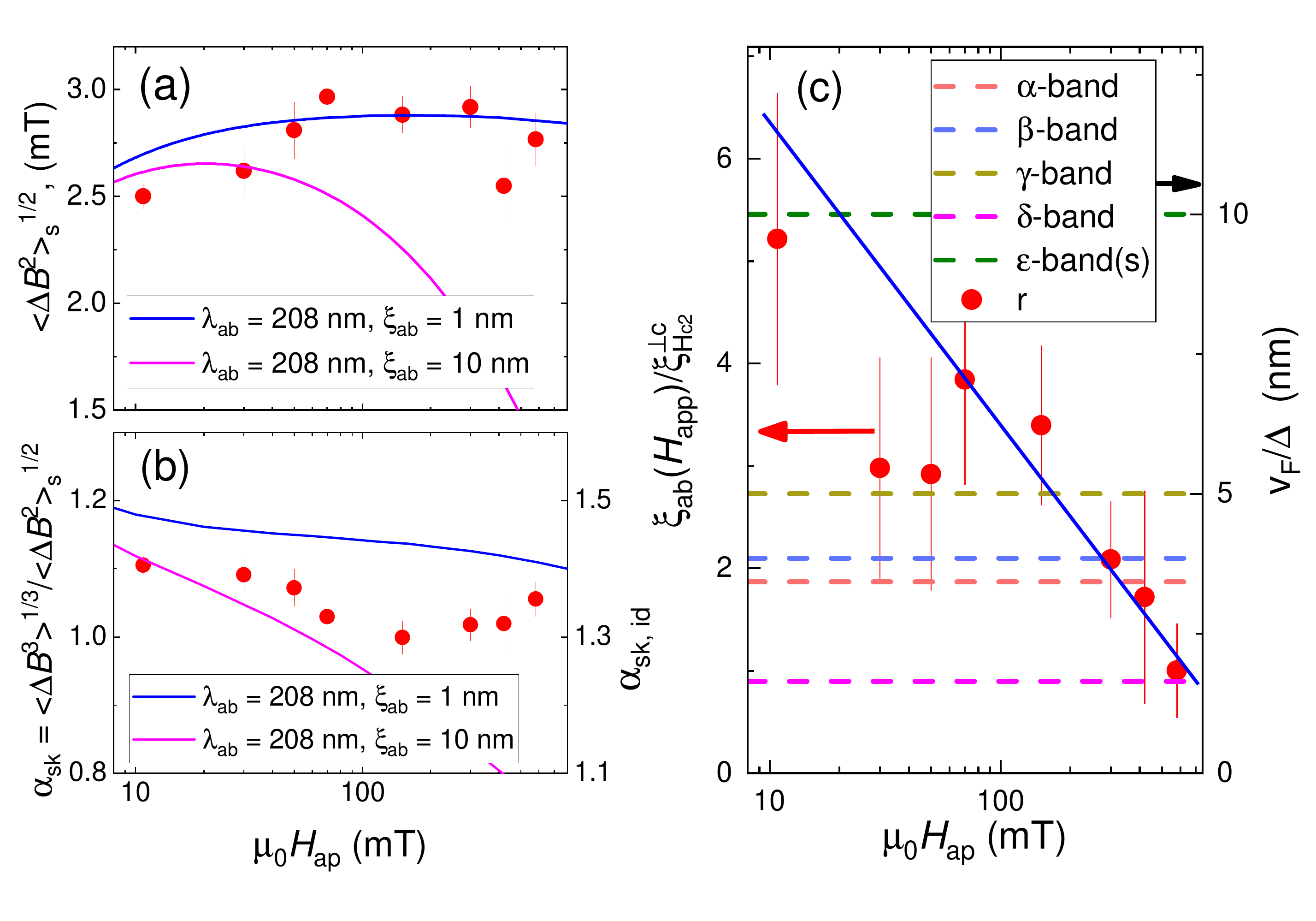}
%
\caption{(a) Field dependence of $\langle \Delta B^2 \rangle ^{1/2}_s$ of CaKFe$_4$As$_4$ at $T=1.55$~K. (b) Dependence of $\alpha_{\rm sk}$ on $H_{\rm ap}$.  The pink and blue solid lines in (a) and (b) are  calculations obtained within the framework of the London model with Gaussian cut-off for $\lambda_{ab}=208$~nm, $\xi_{ab}=10$~nm  and   $\lambda_{ab}=208$~nm, $\xi_{ab}=1$~nm, respectively. (c) Dependence of $\xi_{ab}/ \xi_{\rm H_{c2}}^{\perp {\rm c}}$  on $H_{\rm ap}$. Dashed lines correspond to $\langle v_{\rm F}\rangle_i / \Delta_{0,i}$ values ($i=\alpha$, $\beta$, $\gamma$, $\delta$, or $\epsilon$ is the band index).  }
 \label{fig:SecMom-Alpha}
\end{figure}

The first moment ($\langle B \rangle$), the second- ($\langle \Delta B^2 \rangle$) and third-central moments ($\langle \Delta B^3 \rangle$) of $P(B)_s$ were obtained analytically (see the Supplemental Information, Ref.~\onlinecite{Supplemental_Material}) and have the following physical interpretations: (i) The first moment (the mean field) is generally smaller than $H_{\rm ap}$, due to the diamagnetic nature of the superconducting state. The field shift $\mu_0H_{\rm ap}-\langle B\rangle$ scales with the sample magnetization.\cite{Weber_PRB_1993} (ii) The second moment encodes the broadening of the signal, and contains contributions from the vortex lattice and the nuclear dipole field.  The assumedly temperature independent nuclear dipolar field contribution ($\sigma_{\rm nm}$) is determined from measurements made above $T_{\rm c}$.  The superconducting component ($\langle \Delta B^2 \rangle_s$) is then obtained by subtracting $\sigma_{\rm nm}$ from the measured second moment: $\langle \Delta B^2 \rangle_s =\langle \Delta B^2 \rangle - \sigma_{nm}^2$.  $\langle \Delta B^2 \rangle_s$ is a function of the magnetic penetration depth $\lambda$ and  the vortex core size $\xi\sim \xi_{\rm H_{c2}}$ ($\xi_{\rm H_{c2}}$ is the coherence length as obtained from the upper critical field). In extreme type-II superconductors ($\lambda\gg\xi_{\rm H_{c2}}$) and for fields much smaller than the upper critical field $\langle \Delta B^2 \rangle_s$ is proportional to $\lambda^{-4}$. \cite{Brandt_PRB_1988, Brandt_PRB_2003}
(iii) The third moment accounts for the asymmetric shape of $P(B)$, which is described via the skewness parameter $\alpha_{\rm sk}=\langle\Delta B^3 \rangle^{1/3}/\langle\Delta B^2 \rangle^{1/2}_s$.
In the limit of $\lambda\gg\xi$ and for realistic measurement conditions $\alpha_{\rm sk}\simeq 1.2$ for a well arranged triangular vortex lattice. It is very sensitive to structural changes of the vortex lattice which may occur as a function of temperature and/or magnetic field.\cite{Lee_PRL_1993, Aegerter_PRB_1998}

Figures \ref{fig:SecMom-Alpha}~(a) and (b) show the dependence of the square root of the second moment $\langle \Delta B^2 \rangle^{1/2}_s$  and the skewness parameter $\alpha_{\rm sk}$ on $H_{\rm ap}$ at $T=1.55$~K. Since in our experiments the magnetic field was applied along the crystallographic $c-$axis, both $\langle \Delta B^2 \rangle^{1/2}_s$ and $\alpha_{\rm sk}$ are functions of the in-plane components of the magnetic penetration depth ($\lambda_{ab}$) and the vortex core size ($\xi_{ab}$).
The experimental data were compared with calculations performed within the framework of the London model with Gaussian cut-off for an ideal hexagonal vortex lattice (see the Supplemental part, Ref.~\onlinecite{Supplemental_Material} and Refs.~\onlinecite{Brandt_PRB_1988, Brandt_LTP_1977, Brandt_LTP_1988, Rammer_PhysC_1991}).

The analysis reveals that both experimental $\langle \Delta B^2 \rangle^{1/2}_s$ and $\alpha_{\rm sk}$ field dependencies can be described with an essentially field independent $\lambda_{ab}\simeq 208$~nm and $\xi_{ab}$ ranging from $\simeq 1$ to $10$~nm. As an example, solid curves in Figs.~\ref{fig:SecMom-Alpha} (a) and (b) correspond to the theory calculations for an ideal vortex lattice with $\lambda_{ab}=208$~nm, $\xi_{ab}=10$~nm  (pink curve) and    $\lambda_{ab}=208$~nm, $\xi_{ab}=1$~nm (blue curve). Note that due to vortex lattice distortions, which are always present  in real superconducting samples, the experimentally observed $\alpha_{\rm sk}$ is slightly smaller than the ideal value $\alpha_{\rm sk, id}$ obtained after calculations. \cite{Maisuradze_JPCM_2009}
The independence of $\lambda$ on the magnetic field is a characteristic feature of fully gapped superconductors. As shown by Kadono,\cite{Kadono_JPCM_2004} $\lambda$ increases with increasing field if the superconducting gap contains nodes and is field independent if the gap is isotropic. Our results suggest, therefore, that CaKFe$_4$As$_4$ is a nodeless superconductor in good agreement with the previously reported data.\cite{Mou_PRL_2016,Cho_PRB_2017,Fente_Arxiv_2016,Biswas_PRB_2017}

According to Fente {\it et al.}\cite{Fente_Arxiv_2016}, the  vortex core size in CaKFe$_4$As$_4$ {\it decreases} with increasing field. Our experimental data are consistent with this finding. Figures~\ref{fig:SecMom-Alpha}~(a) and (b) imply that the low-field $\langle B^2\rangle^{1/2}_s$ and $\alpha_{\rm sk}$ points stay closer to the $\xi_{ab}=10$~nm theory curves, while the high field values are closer to $\xi_{ab}=1$~nm curves. The field-induced decrease of $\xi_{ab}$ also accounts for the local minimum on $\alpha_{\rm sk}(H_{\rm ap})$ at $\mu_0 H_{\rm ap}\simeq 150$~mT [see Fig.~\ref{fig:SecMom-Alpha}~[b]). Under the assumption of field independent $\lambda_{ab}=208(4)$~nm and $\xi_{ab}(0.58\ {\rm T})=\xi_{\rm H_{c2}}^{\perp {\rm c}}$ ($\xi_{\rm H_{c2}}^{\perp {\rm c}}=1.43$~nm is the coherence length as obtained from the upper critical field, Ref.~\onlinecite{Meier_PRB_2016}) the field dependence of $\xi_{ab}(H_{\rm ap})/\xi_{\rm H_{c2}}^{\perp {}\rm c}$ was reconstructed [see Fig.~\ref{fig:SecMom-Alpha}~(c) and the Supplemental Information for details, Ref.~\onlinecite{Supplemental_Material}].

\begin{table}[htb]
\caption{\label{tab1} Parameters extracted and calculated from ARPES and $\mu$SR data. $\Delta_{0,i}$ is the zero-temperature value of the superconducting gap, $\langle v_{\rm F}\rangle_i$ is the mean value of the Fermi velocity,  $\langle d_{\rm F}\rangle_i$ is the average diameter of the Fermi surface sheet, and $\lambda_{ab,i}^{-2}(0)/\lambda_{ab}^{-2}(0)$ is the relative contribution of $i-$th band to $\lambda_{ab}^{-2}(0)$.}
\begin{tabular}{l|ccccc}
\hline
\hline
                   &$\Delta_{0,i}$&$\langle v_{\rm F}\rangle_i$& $\langle d_{\rm F}\rangle_i$& $\frac{\lambda_{ab,i}^{-2}(0)}{\lambda_{ab}^{-2}(0)}$& Technique\\
                   &(meV)      & (eV$\cdot {\rm nm}$)  & (${\rm nm}^{-1}$)      & &\\
\hline
$\alpha-$band & 10.5(0.8)   & 0.036(1)                 & 1.60(25)  &0.073(11)& \multirow{4}{*}{ARPES}\\
$\beta-$band  & 13.0(0.8)   & 0.050(2)                 & 3.23(25)  &0.202(15)&\\
$\gamma-$band & 8.0(0.6)    & 0.040(1)\footnotemark[1] & 6.53(25)  &0.323(12)&\\
$\delta-$band & 12.0(1.3)   & $\sim0.008$                               & 3.30(25)  &0.032(3)&\\
\hline
$\epsilon-$bands& 2.4(2)   & $\sim 0.025$& $\sim10.0$\footnotemark[2] & 0.370(40)&$\mu$SR\\
\hline
\hline
\end{tabular}
 \footnotetext[1]{Averaged over $h\nu=6.7$ and 21.2~eV photon energies, Fig.~\ref{fig:ARPES}.}
 \footnotetext[2]{The sum of diameters.}
\end{table}

\begin{figure}[htb]
\includegraphics[width=1.0\linewidth]{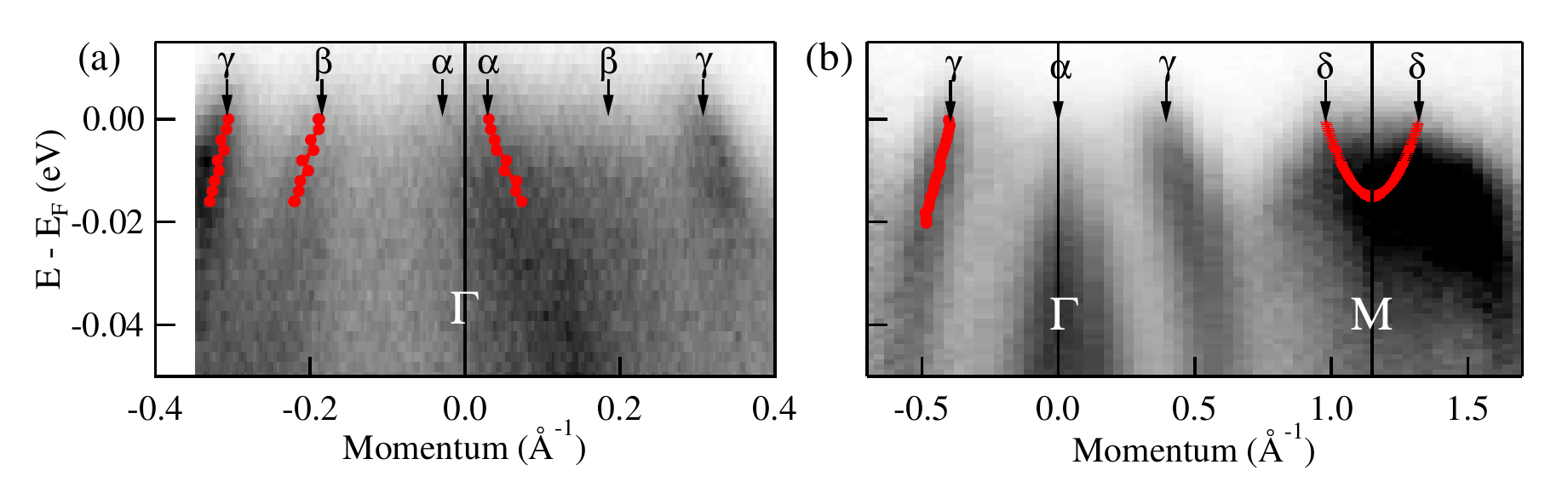}
%
\caption{Band dispersions of CaKFe$_4$As$_4$ at $T=40$~K obtained in ARPES experiments: (a) by using the laser light source at the photon energy $h\nu=6.7$~eV; (b) by using plasma Helium lamp with $h\nu=21.2$~eV. Red lines are fits of the band dispersion curves near zero-energy  allowing to obtain the Fermi velocities. The units of momentum are kept in ${\rm \AA}^{-1}$ for consistency with ARPES works. }
 \label{fig:ARPES}
\end{figure}

The decrease of the vortex core size with increasing field was observed in various conventional and unconventional superconductors in tunneling, magnetisation and $\mu$SR experiments (see {\it e.g.} Refs.~\onlinecite{Sonier_RMP_2000, Sonier_JPCM_2004, Kogan_PRB_2006, Fente_PRB_2016}) as well as reported for CaKFe$_4$As$_4$ in tunneling experiments for fields exceeding 0.5~T.\cite{Fente_Arxiv_2016} The strongest effect was observed in multigap superconductors like MgB$_2$, NbSe$_2$ \cite{Sonier_JPCM_2004,Eskildsen_PRL_2002} and it was explained by the gap and Fermi velocity dependent length scales $\xi_i\propto \langle v_{\rm F}\rangle_i/\Delta_{0,i}$ ($i$ is the band index and $ \langle v_{\rm F}\rangle$ is the mean value of the Fermi velocity). At low and high magnetic fields the vortex core size is governed by the high and the low $\langle v_{\rm F}\rangle_i/\Delta_{0,i}$ ratios, respectively.
Previous ARPES experiments on CaKFe$_4$As$_4$ reveal the presence of at least four bands crossing the Fermi level with the corresponding $\Delta_0$ values  summarized in Table~\ref{tab1}.\cite{Mou_PRL_2016} Fermi velocities were further obtained in this work by performing linear fits of the band dispersion curves near the Fermi energy (see Fig.~\ref{fig:ARPES} and Table~\ref{tab1}). Obviously, the four bands reported in ARPES experiments do not explain the $\xi_{ab}/\xi_{\rm H_{c2}}^{\perp {\rm c}}$ {\it vs.} $H_{\rm ap}$ behavior shown in Fig.~\ref{fig:SecMom-Alpha}~(c). One needs to assume the presence of an additional band, or series of bands ($\epsilon-$bands), with $\langle v_{\rm F}\rangle_\epsilon/\Delta_{0,\epsilon}\simeq$~10~nm. Bearing in mind that a small gap of $\Delta_{0,\epsilon}\simeq$~2.4~meV was obtained in our $\lambda_{ab}^{-2}(T)$ studies (see the following discussion), as well as reported in tunneling and superfluid density experiments by other groups,\cite{Cho_PRB_2017,Fente_Arxiv_2016,Biswas_PRB_2017} the average Fermi velocity within the $\epsilon-$bands is $\langle v_{\rm F}\rangle_\epsilon \simeq$~0.025~eV$\cdot {\rm nm}$. This value is well within the range for $\langle v_{\rm F}\rangle$'s obtained by ARPES (see Table~\ref{tab1}).

\begin{figure}[htb]
\includegraphics[width=1.0\linewidth]{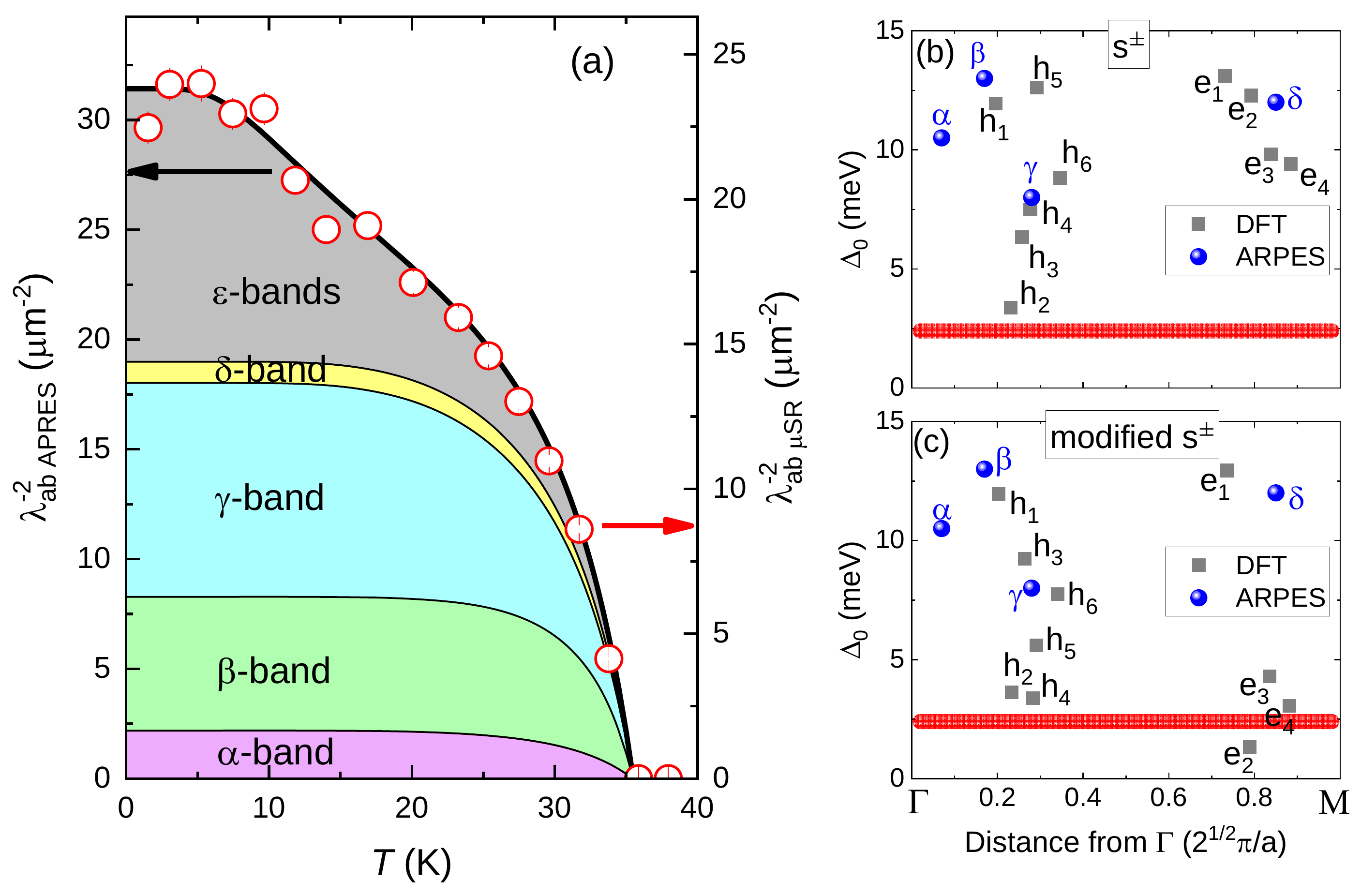}
%
\caption{(a) $\lambda_{ab}^{-2}(T)$  obtained from the measured $\langle B^2\rangle^{1/2}$ at $\mu_0 H=11$~mT ($\lambda^{-2}_{ab, \mu{\rm SR}}$) and inferred from ARPES data ($\lambda^{-2}_{ab,{\rm ARPES}}$). The colored stripes are contributions of various Fermi surfaces to $\lambda_{ab}^{-2}(T)$ . (b) Gaps as a function of momentum distance from $\Gamma$ in hole ($h$) and electron ($e$) pockets as obtained from DFT calculations within the framework of the $s^\pm-$model.\cite{Lochner_PRB_2017} (c) the same as in (b) but for a modified $s^\pm-$model. The red line is $\Delta_0=$2.4~meV. The blue points refer to the ARPES values.\cite{Mou_PRL_2016}  }
 \label{fig:Lambda2}
\end{figure}

Fig.~\ref{fig:Lambda2}~(a) compares the $\lambda_{ab}^{-2}(T)$ dependence obtained in $\mu$SR experiment from the measured $\langle B^2\rangle^{1/2}$ at $\mu_0 H=11$~mT (red open circles) with the one calculated from the electronic band dispersion and the momentum resolved superconducting gap measured by ARPES. Following Refs.~\onlinecite{Evtushinsky_NJP_2009, Tinkham_book_75, Khasanov_PRL_2009, Khasanov_PRL_La214_07}, $\lambda_{ab}^{-2}(T)$ is determined as:
\begin{equation}
\lambda_{ab}^{-2}(T) = \sum_i I_i \left[ 1 +2\int_{\Delta_i(T)}^{\infty}\left(\frac{\partial
f}{\partial E}\right)\frac{E dE}{\sqrt{E^2-\Delta_i(T)^2}} \right].
\end{equation}
Here $f=[1+\exp(E/k_BT)]^{-1}$ is  the Fermi function, $\Delta_i(T)=\Delta_{0,i} \tanh\{1.82[1.018(T_{\rm c}/T-1)]^{0.51}\} $,\cite{Carrington_03} and $I_i$ is the contribution of the $i-$th gap to $\lambda_{ab}^{-2}(T=0)$, which is obtained as:\cite{Evtushinsky_NJP_2009}
\begin{equation}
I_i = \frac{e^2}{2\pi \varepsilon_0 c^2 h L_c}  \oint_{i-{\rm th \ band}} v_{{\rm F},i}({\bf k})
{\rm d}k.
\label{eq:Lambda0}
\end{equation}
Here $e$ is the elementary charge, $\varepsilon_0$ is the electric constant, $h$ is the Planck constant, $c$ is the speed of light, $L_c$ is the $c-$axis lattice parameter and ${\bf k}$ is a momentum vector within the reciprocal space. Integrations are made over the corresponding Fermi surface contours.

The analysis reveals that the four bands observed in the ARPES experiment alone do not describe the experimentally measured $\lambda_{ab}^{-2}(T)$. In analogy with the above discussed $\xi_{ab}(H_{\rm ap})$ results [see Fig.~\ref{fig:SecMom-Alpha}~(c)] the presence of  additional bands with smaller superconducting energy gaps are needed. The difference between the calculated and measured $\lambda_{ab}^{-2}$ allows access to the zero-temperature value of the superconducting gap of the  $\epsilon-$bands. A reasonably good agreement is achieved for $\Delta_{0,\epsilon}\simeq$~2.4(2)~meV, $\lambda_{ab, {\rm ARPES}}(0)\simeq$~187(11)~nm and a contribution to the zero-temperature superfluid density [$\lambda_{ab,\epsilon}^{-2}(0)/\lambda_{ab}^{-2}(0)$] of $\simeq 37(4)$\% [see Fig.~\ref{fig:Lambda2}~(a),  Table ~\ref{tab1} and the Supplemental Information for details, Ref.~\onlinecite{Supplemental_Material}].  With the known value of the Fermi velocity $\langle v_{\rm F}\rangle_\epsilon\simeq$~0.025~eV$\cdot {\rm nm}$ (see Table~\ref{tab1}), a sum of the diameters of  additional bands $\langle d_{\rm F}\rangle_\epsilon\simeq$~10.0~${\rm nm}^{-1}$ was found.
Two important points need to be mentioned.
(i) The data in Fig.~4a were analyzed using only three independent parameters [$\Delta_{0,\epsilon}$, $\langle d_{\rm F}\rangle_\epsilon$ and $n=\lambda_{ab, {\rm ARPES}}(0) / \lambda_{ab, \mu {\rm SR}}(0) $]. The rest of the parameters were fixed to the values obtained in ARPES, magnetization and  $\xi_{ab}(H_{\rm ap})$ measurements [see Fig.~2~(c),  Ref.~\onlinecite{Meier_PRB_2016} and Table~\ref{tab1}].
(ii) The zero temperature value of the in-plane penetration depth calculated from the ARPES data (including  contribution of $\epsilon-$bands) results in $\lambda_{ab, {\rm ARPES}}(0)\simeq$~187(11)~nm, which is approximately 10\% lower than $208(4)$~nm as determined by the $\mu$SR experiment. Note that a similar difference was reported for Ba$_{1-x}$K$_x$Fe$_2$As$_2$ in Refs.~\onlinecite{Evtushinsky_NJP_2009, Khasanov_PRL_2009}.

Temperature dependencies of the superfluid density components of $\alpha-$, $\beta-$, $\gamma-$ and $\delta-$bands follow the BCS type of mean-field behaviour, while the response of $\varepsilon-$bands is markedly different (see Fig.~3 in the Supplemental Information, Ref.~\onlinecite{Supplemental_Material}). This may suggest that $\varepsilon-$bands are only weakly involved into superconductivity, and alone would have $T_c$ of the order of 15~K only. Above 15~K the superconductivity in $\varepsilon-$bands remains due to effects of interband coupling.\cite{Suhl_PRL_1959, Moskalenko_PMM_1959}

The $s^\pm$ gap symmetry, with a sign change between hole and electron pockets, in CaKFe$_4$As$_4$ was supported by the observation of the  neutron-spin resonance peak with a characteristic energy $\simeq 12.5$~meV at the antiferromagnetic wave vector ${\bf Q}_{\rm AF}= (\pi,\pi)$.\cite{Iida_JPCJ_2017,comment} We should emphasize, however that the recent DFT calculations obtain two stable solutions for possible gap functions in CaKFe$_4$As$_4$. The first corresponds to the conventional $s^\pm-$state, while the second (modified  $s^\pm-$state) allows an additional sign change within the hole and electron pockets with some gaps being small in magnitude.\cite{Lochner_PRB_2017} The modified $s^\pm-$state enhances the spin response at the antiferromagnetic wave vector $(\pi,\pi)$ and, therefore, becomes consistent with the results of Iida {\it et al.}\cite{Iida_JPCJ_2017} This enhancement, however, cannot be regarded as a pure spin resonance due to relatively small gap sizes on some electron and hole pocket bands. As a consequence, the resonance in CaKFe$_4$As$_4$ cannot be treated as a true spin excitation, since its position is not {\it below} but {\it inside} the particle-hole continuum (see the Supplemental Information for details, Ref.~\onlinecite{Supplemental_Material}).

Figures~\ref{fig:Lambda2}~(b) and (c) show the superconducting gaps in hole ($h$) and electron ($e$) pockets as a function of momentum distance from the $\Gamma$ point (the center of the Brillouin zone) as obtained from DFT calculations within the framework of the $s^\pm$ and modified  $s^\pm$ models, respectively.\cite{Lochner_PRB_2017} Within the $s^\pm$ approach only one out of ten bands has the gap value smaller than 5~meV [Fig.~\ref{fig:Lambda2}~(b)]. Note, that due to the so-called red/blue shift, reported by the authors of Ref.~\onlinecite{Lochner_PRB_2017}, one can not rely on the band diameters obtained from the DFT calculations, so the analysis similar to the one made above for obtaining $\lambda_{ab,{\rm ARPES}}^{-2}(T)$ can not be performed. One would expect, however, that the contribution of 9 bands to $\lambda_{ab}^{-2}(T)$ should be at least twice as high as for 4 bands obtained in ARPES experiments [see Fig.~\ref{fig:Lambda2}~(a)]. This would lead to complete disagreement between the calculated and experimentally measured $\lambda_{ab}^{-2}(T)$ dependencies.  In contrast, within the modified  $s^\pm-$model 5 (2 hole and 3 electron) bands have gap values close to $\Delta_0=$~2.4~meV [Fig.~\ref{fig:Lambda2}~(c)]. They may account, therefore, for 37\% contribution to $\lambda_{ab}^{-2}(0)$ (see Table~\ref{tab1}). The other 5 bands left (4 hole and 1 electron band) result in gap values comparable with that obtained in ARPES experiments.
This implies that the modified  $s^\pm-$ model agrees better with our experimental findings.

To conclude, the temperature and the magnetic field dependencies of the in-plane magnetic penetration depth $\lambda_{ab}$ and the field dependence of the vortex core size $\xi_{ab}$ in CaKFe$_4$As$_4$ single crystal were studied by means of muon-spin rotation. A comparison of the temperature dependence of $\lambda_{ab}^{-2}$ measured by $\mu$SR to the one determined from ARPES experiments confirms the presence of multiple gaps at the Fermi level. An agreement between the $\mu$SR and ARPES data requires the presence of  additional bands which are characterised by small superconducting gaps with an averaged zero-temperature value of $\Delta_{0} \simeq$~2.4(2)~meV.
Our data suggest that the order parameter in CaKFe$_4$As$_4$ may acquire a more complex form than the simple $s^\pm-$symmetry. The gap sign change occurs  not just between the electron and hole pockets but also within pockets of similar type. Such a gap state is favored by weakened interband repulsion and yields further interesting consequences for the spin resonance, which needs to be further verified experimentally

The work was performed at the Swiss Muon Source (S$\mu$S), Paul Scherrer Institute (PSI, Switzerland).
RK acknowledges  D.V.~Evtushinsky for helpful discussions and J. Barker for help with the manuscript preparation. WRM was funded by the Gordon and Betty Moore Foundation EPiQS Initiative through Grant GBMF4411. Work at Ames Laboratory was supported by the U.S. Department of Energy, Office of Science, Basic Energy Sciences, Materials Science and Engineering Division. Ames Laboratory is operated for the U.S. DOE by Iowa State University under Contract No. DE-AC02-07CH11358.

\end{document}